\documentclass[aps,twocolumn,showpacs,floatfix]{revtex4}
\usepackage[ansinew]{inputenc}
\usepackage{amssymb}
\usepackage{graphicx}
\usepackage{amsmath}

\begin{document}
\title{Quantum superposition and entanglement of mesoscopic plasmons}
\author{Sylvain Fasel}
\email{sylvain.fasel@physics.unige.ch}
\author{Matthäus Halder}
\author{Nicolas Gisin}
\author{Hugo Zbinden}
\affiliation{Group of Applied Physics, University of Geneva, CH-1211 Geneva 4, Switzerland}
\begin{abstract}
Quantum superpositions and entanglement are at the heart of the   
quantum information science. There have been only a few investigations of these phenomena at the mesoscopic level, despite the fact that these systems are promising for quantum state storage and processing.
Here we present two novel experiments with surface plasmons propagating  
on cm--long metallic stripe waveguides. We demonstrate that two plasmons can be entangled at remote places.
In addition, we create a single plasmon in a temporal superposition state:  
it exists in a superposition of two widely separated moments. These quantum  
states, created using photons at telecom wavelength, are collectively held  
by a mesoscopic number of electrons coding a single quantum bit of  
information; They are shown to be very robust against decoherence.
\end{abstract}

\pacs{03.67.Mn, 42.50.Dv, 71.36.+c, 73.20.Mf, 78.66.Bz}

\maketitle

Quantum superpositions and entanglement are widely
recognized as the core of quantum physics, with all its
counter-intuitive features. They are also essential for the coming
age of quantum technology, as needed for instance for quantum
information processing. Entanglement of several photons
\cite{zeilingerGHZ,pan5photons,qutrits,zeilingerdanube,swapping} or ions \cite{monroe4ions,blattcnot,monroe2,winelandteleport,blattteleport} is nowadays
common in laboratories, but not much is known at the mesoscopic
level. Surface plasmons (SPs) are propagating charge density
waves, involving about $10^{10}$ free electrons at the surface of
metals \cite{plasmonsbook,plasmonsreview}. SPs can be excited using light, and are thus good
candidates to test the robustness of quantum superpositions and entanglement at a mesoscopic level. {Previous experiments focused on the conservation of entanglement for photon--plasmon--photon conversion using metallic sub--wavelength hole--arrays \cite{polentang} and long range SPs device \cite{plasmonsfasel}}.
In this letter we use cm--long plasmon guides to experimentally investigate these aspects of quantum physics. To begin, we demonstrate for the first time entanglement between two remote SPs {existing at the same time and without photons being present simultaneously}. In a second experiment, we create temporal superposition states at extreme scales. SPs have a finite life-time: they are created, propagate and eventually die.
We create SPs with photons in superposition of two
time-bins and consequently this plasmonic processes is in a coherent
superposition of occurring at two times that differ by much more
than its life--time. At a macroscopic level, this would superpose a cat
living at two epochs that differ by much more than a cat's
life--time.
One should stress that the investigated system involves a mesoscopic number of electrons coding a single quantum degree of
freedom; hence they are not proper Schrödinger cat states. 

\begin{figure}[h]
\includegraphics[width=0.7\columnwidth]{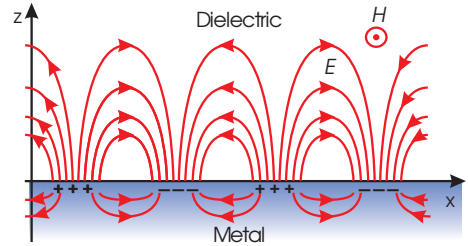}
\caption{Schematic structure of the electric field $E$ of a SP propagating on a metallic surface in the $x$ direction. The field strength decreases exponentially with distance $z$ from the surface. $H$: direction of the magnetic field; + and - indicate respectively area with lower and higher density of electrons. (figure inspired by \cite{plasmonsreview})} \label{SPfield}
\end{figure}

SPs correspond to the propagation of an
electromagnetic quantum field, but with major differences with
respect to the propagation of photons in a fiber. The structure of
the electric field is indeed very different (see figure \ref{SPfield}) in the two cases. Moreover, the implication of the free
electron plasma at the surface of the metal is essential for the
generation of a SP state, {and its quantum properties are completely defined by the geometrical and electro-magnetical characteristics of the solid state of matter that support them.} Therefore the existence of a SPs is {a 
 truly mesoscopic} phenomenon and is very bound to matter, in contrast to the propagation of
photons.

Long range SPs (LR-SPs) are symmetric low loss propagating modes,
that can be excited on thin metallic stripes sandwiched between
dielectrics. A particular geometry of these structures allows the
direct in and out coupling of the 1550\,nm light mode of a
standard telecom fiber to the LR-SPs mode, with very low insertion
losses \cite{lrspp}. These devices are called plasmonic
stripe waveguides (PSW).

{In the first experiment presented in this paper, we focus on the entanglement of distant SPs that fulfill the following two criteria. First, the two SPs being transient processes, care must be taken to excite them at the same time within the SPs life--time. Second, the SPs propagation distance must be larger than the coherence length of the exciting photons, in order to clearly destroy the photon during the process. Hole--arrays based experiments did not met these requirements, but this is achieved in our experiment using LR-SPs.}

\begin{figure}[h]
\includegraphics[width=\columnwidth]{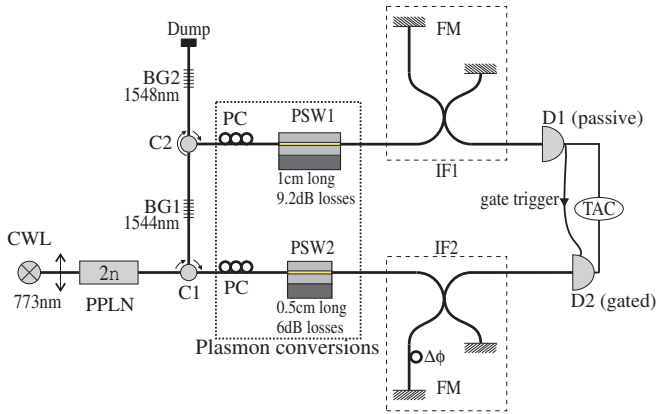}
\caption{Scheme of the experimental setup used for entangling two
SPs on the plasmonic stripe waveguides PSW1 and PSW2.
CWL: continuous wave laser at 773\,nm; C1, C2: circulators; BG1,BG2: tunable Bragg filters; PC:
polarization controller; FM: Faraday mirrors; IF1, IF2: unbalanced Michelson
interferometers; TAC: Time to analog converter; D1, D2: InGaAs avalanche photodiode single photon detectors cooled at -45° C; $\Delta\varphi$: controllable phase shifter. D1 is operated in passive mode (passively quenched by a resistance of 1.3\,M$\Omega$) at 7\% quantum efficiency with 5kHz of noise and a signal of 20kHz. D1 triggers D2, working in gated mode at 15\% efficiency. A successful detection of the corresponding photon leads to a mean coincidence rate of 12 per second. The dead--time of the detection system is about $10\,\mu$s.} \label{schema2p}
\end{figure}

We use two distant similar PSWs inserted in between an energy-time entangled photon pair source and two interferometers, to create and verify the entanglement of SP pairs. The corresponding setup is presented in figure \ref{schema2p}.

The source is a spontaneous parametric down conversion (SPDC) source
consisting of a periodically poled lithium niobate (PPLN)
waveguide {(HC Photonics)} pumped with a continuous-wave laser diode. The phase
matching conditions of the process are such that it takes place at
the degeneracy point, i.e. the two photons are created around the
same average central wavelength of about 1546\,nm. Their spectral
width is about 80\,nm. The photon pairs and the remaining pump
power are butt coupled into a single mode fiber. By energy conservation, (and
provided that the spectrum of the pump laser is of negligible
width), whenever one photon is measured at a given wavelength,
its twin photon can only be detected at a wavelength such that the sum of both energies are equal to the pump photons energy. We use this property to separate the paired photons with high efficiency and
controllability. This is done with tunable fibred Bragg gratings {(AOS GmbH)}
and circulators.  Photons coming from the source with a first chosen wavelength are reflected by the first Bragg grating BG1 and
launched into PSW1 via a first circulator C1. The remaining
light passes through this filter and reaches the Bragg grating
BG2. This filter is tuned such that only the photons that are energetically complementary of the first ones are in turn back-reflected, and then launched into the second PSW via circulator C2. This setup ensures that
only paired photons are injected into the PSWs. It also enables
us to filter out the remaining pump light that is directed to a
dump in order to avoid back reflections. The Bragg
grating filters are tuned by stretching the fiber in
which the grating is inscribed. The spectral width of these
filters is 0.8\,nm, thus the coherence length of the filtered
photons is 0.9\,mm (coherence time of 4.25\,ps) {, i.e. much smaller than the PSWs}.
The optical lengths of the two paths from the source to PSW1 and
PSW2 are equalized with less than 1\,mm uncertainty using optical frequency domain reflectometry
(OFDR) \cite{OFDR}. {The uncertainty on the path lengths from the source to the PSWs is small compared to the length of the shorter PSW (5\,mm). The simultaneous existence of both SPs is thus ensured since SPs on PSWs are not faster than photons in fibers.}

Photons are collected back at the output of the PSWs into single
mode fibers and sent to two matched unbalanced fiber Michelson
interferometers having a path length difference between the two
arms corresponding to 1.2\,ns. They are stabilized in temperature
and the relative phase between the two arms of IF2 can be scanned
by a piezoelectric actuator.

\begin{figure}[h]
\includegraphics[width=\columnwidth]{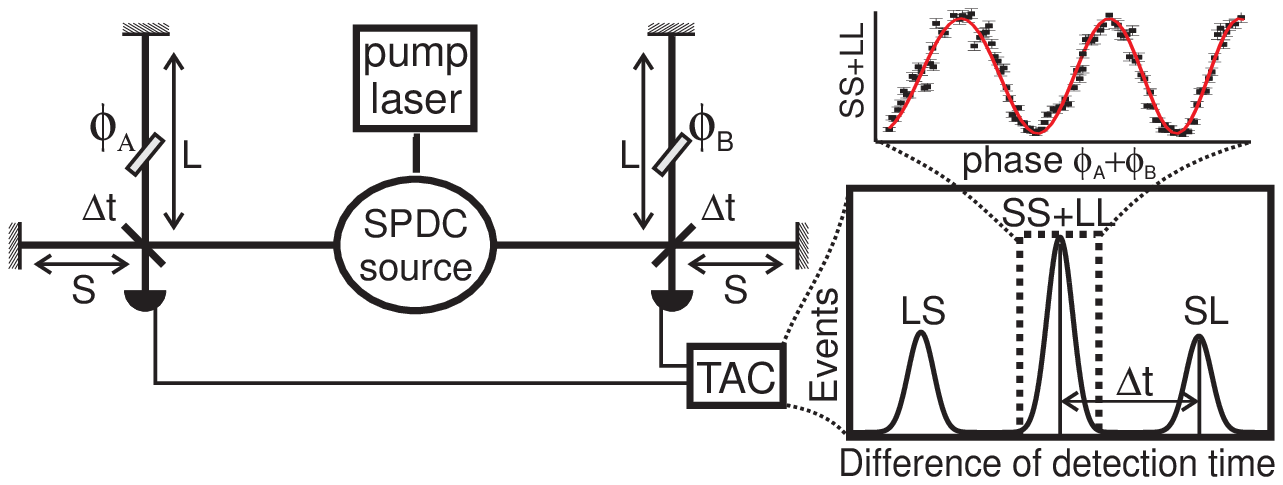}
\caption{Franson-type interferometers-source arrangement. S: short path; L: long path; $\Delta t$: time difference between photons propagations thought long and short arm; $\varphi_{A,B}$: phases applied on photons. Two photons are emitted simultaneously, and independently travel through long or short arms of the interferometers. Only three distinguishable two-photon detection events are possible. Indeed, when photons both choose short or long arms, the two detections occur with the same time differences, and are thus quantum mechanically undistinguishable, under the condition that the pump laser coherence length is larger than the L-S path difference. The detection rate corresponding to these events is discriminated using a time-window. This rate is modulated by quantum interferences as a function of the sum of the phases applied in the interferometers, and is recorded as sinusoidal fringes. These are the signature of the energy-time entanglement, and their visibility is directly related to the degree of entanglement.} \label{franson}
\end{figure}

InGaAs avalanche photodiode (APD) detectors are connected at the output of each interferometer. Note that the first InGaAs APD (D1) {(Epitaxx)} is operated in passive mode \cite{HalderPassiv,RarityPassiv} and its detection signal triggers the detection gate of the second (D2), which is operated in gated mode {(IdQuantique)}. In order to maximize the signal-to-noise ratio, which means maximizing the probability of getting a corresponding count at D2
conditioned on a photon detection at D1, the more lossy path is connected to the passive APD D1. The difference of the detection times is recorded using a time to amplitude converter (TAC). Using this setup, we performed a conventional Franson-type entanglement measurement \cite{franson} (details on figure \ref{franson}).

\begin{figure}[h]
\includegraphics[width=0.9\columnwidth]{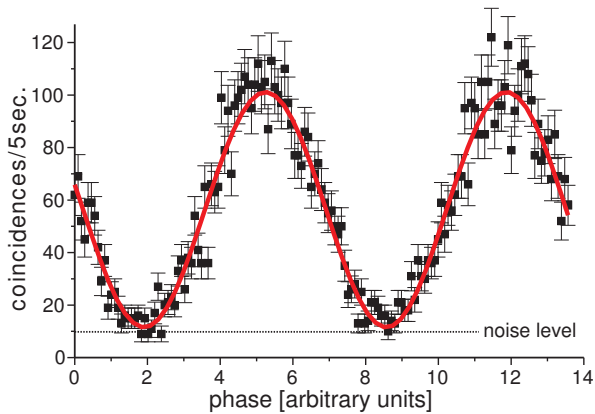}
\caption{Interference fringes measured with the two PSWs in the
path of the photons. The visibility, obtained trough sinusoidal
fitting (solid curve), is $96.5\pm1.6\%$, while the visibility of
the reference fringes measured without PSWs (fringes not shown) is
$97.4\pm1.2\%$.} \label{results2p}
\end{figure}

The result of this measurement is presented in figure \ref{results2p}. In the same way we measured the entanglement of the original photon pairs emitted by the source as a reference. The interference fringes
recorded after photon--plasmon--photon conversion and the reference exhibit the same
high visibility (inside the error uncertainty) of about 97\% in both cases.

{This result demonstrates that the degree of entanglement is not lowered with respect to the initial two-photon entanglement. At some point the entanglement is therefore entirely carried by two SPs since the experimental conditions fulfill the required criteria. To our knowledge, this is achieved for the first time. We demonstrated the creation of entangled SP pairs, and hence the entanglement of two distant mesoscopic systems (separated by about 1\,m) constituted by the very large number of free electrons at the surface of the two PSWs. These systems, although consisting of many particles, could code entangled qubit (e.g.  time-bin qubits \cite{timebin}) in a simpler way than other experiments relying on different type of plasmonic devices \cite{polentang} or atomic ensembles \cite{polziknature,polzik3,collectivatoms}. The present setup is working at standard telecom wavelength and with a form of entanglement which is robust for transmission in fibers \cite{robust}.}

\begin{figure}[h]
\includegraphics[width=\columnwidth]{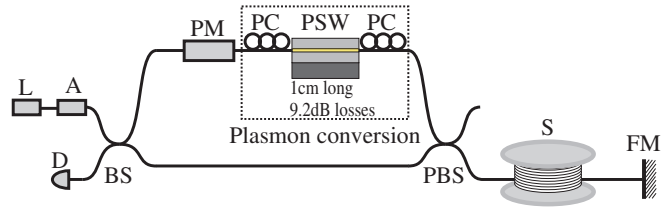}
\caption{Scheme of the experimental setup for the coherent
superposition of plasmons at two instants of existence. All fibers
are polarization maintaining, except the fiber spool and the fibers of the plasmon conversion
part. L: pulsed laser at
1550nm, repetition rate: 5\,Mhz, pulse length 1.2\,ns; A: variable
attenuator; D: InGaAs peltier cooled photodiode single photon counter;
BS: 50/50 beam splitter; PM: phase modulator; PC: polarization
controller; PSW: surface plasmons stripe waveguide; PBS:
polarization beam splitter; S: standard fiber spool of several kilometers; FM: Faraday mirror} \label{pnp}
\end{figure}

{In the second experiment we create a single SP
in a coherent superposition of two widely separated instants of existence. We verify whether the coherence is preserved even
if the temporal separation of the two instants of existence is
several orders of magnitude larger than the SP life--time.} For this purpose, we used the auto--compensating interferometer described in figure \ref{pnp}, consisting of an unbalanced
Mach-Zehnder interferometer connected by a spool of fiber to a Faraday mirror.

The time scale for a change in the length of the
interferometer due to temperature variation is slower than the time needed for the two-way travel of the light, {and the whole system behaves like a single very large and symmetric interferometer }(when applied to quantum cryptography, this configuration is known as the "Plug'\&'Play" system) \cite{pnp}. A PSW of 1\,cm length is placed in the long arm of the Mach-Zehnder interferometer. 
{Pulses can propagate following two paths of equal lengths. One corresponds to pulses that first choose the Mach-Zehnder long arm, undergo photon--plasmon--photon conversion on the PSW (and thus achieve the full SP creation, propagation, and recollection), travel back and forth through the fiber spool, and then take the short arm of the Mach-Zehnder. The other path corresponds to pulses that first choose short arm, travel back and forth, and then excite a SP in the long arm. The SP conversion does thus not occur at the same time for the two paths, but at instants separated by the time needed for pulses to travel twice inside the fiber spool (i.e. two times 5\,$\mu$s multiplied by the length of the fiber in kilometer). Pulses are finally detected at one output of the interferometer.

The two paths are undistinguishable and we are thus in presence of a coherent superposition, but only as long as the PSW does not introduce distinguishability between them. At this condition only, one can observe interference at the detector, and the detection probability is a sinusoidal function of the phase shift which is applied in synchronization with the returning pulses. These interference fringes are recorded by sending several light pulses and applying different phase values. The visibility of these fringes represents a direct indication of the coherence of the created superposition state for the path, and thus for the existence time of the plasmonic processes.}

\begin{figure}[h]
\includegraphics[width=0.9\columnwidth]{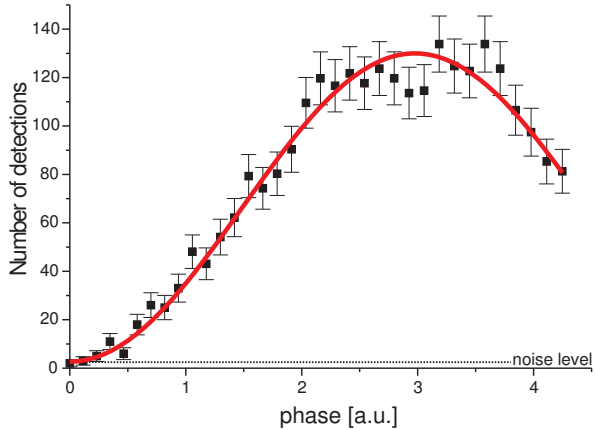} 
\caption{Interference fringes measured with the PSW in the path of the photons. The average number of photons per pulses was 1, and the delay $\Delta t$ was about 270\,$\mu$s (this correspond to a fiber length of 27\,km). The visibility, obtained through sinusoidal fitting (solid curve), is $99.4\pm1.1\%$ in this case, compatible with the visibility of the reference fringes measured without PSW (fringes not shown).}
\label{resultspnp}
\end{figure}

We recorded interference fringes with and without PSW in the path of the photons. We first recorded fringes with an average number of photons per pulse set to 1. More precisely, this value is adjusted (using the variable attenuator) so that the sum of the average number of photons in the short arm and in the long arm just before the PSW is 1. The results of this measurement for a short delay $\Delta t$ is presented on figure \ref{resultspnp}. For larger delays, we increased the launched pulse energy in order to obtain good statistics on the detection counts. 

We repeated the experiment many times for various pulse powers and for several different delays ranging from 0.27\,ms to 1.24\,ms (corresponding to spooled fiber length from 27\,km to 124\,km). In every cases, we consistently found visibilities higher than 99\%. The time needed for a SP to propagate from the input to the output of the PSW (i.e. the "life--time" of the SP), is of the order of 50\,ps. The maximal delay we used in our experiment (1.24\,ms) is therefore more than $10^7$ times larger. We thus demonstrated that SPs can be in a coherent superposition state of existing at two times separated by a large delay, even if this delay is much larger than their life--time.
    
The two presented experiments demonstrate that quantum
superpositions and entanglement can be surprisingly robust. This
adds to the growing experimental evidence that robust manipulation of
entanglement is feasible
\cite{zeilingerGHZ,pan5photons,qutrits,zeilingerdanube,swapping,monroe4ions,blattcnot,monroe2,winelandteleport,blattteleport} with today's technology. It stresses that quantum bits can be carried by
collective modes of a mesoscopic number of particles, here
electrons. However, one should emphasize that in the reported
experiments, as in similar ones \cite{polentang,polziknature,polzik3,collectivatoms}, the many particles
collectively code for only a very limited number of degrees of
freedom. These results are thus not in conflict with the well
established theory of decoherence. Entangling many degrees of
freedom, or equivalently many quantum bits, remains a challenge, however
the present results are encouraging.

\begin{acknowledgments}
We would like to thank Peter W. Madsen and Thomas Nikolajsen from Micro Managed Photons A/S (Denmark) for providing PSW samples and expert advices, and Franck Robin, Daniel Erni and Esteban Moreno for stimulating discussions. Financial support by the Swiss
NCCR Quantum Photonics is acknowledged.
\end{acknowledgments}

\end{document}